\documentclass[aps,pra,twocolumn,groupedaddress,showpacs]{revtex4}
\usepackage{amssymb}
\usepackage{amsmath}
\usepackage{dcolumn}
\usepackage{bm}
\usepackage{graphicx}
\usepackage{mathrsfs}
\usepackage{color}

\begin{document}

\title{Method for observing robust and tunable phonon blockade \\in a
nanomechanical resonator coupled to a charge qubit}
\author{Xin Wang$^{1,2}$}
\author{Adam Miranowicz$^{3,2}$}
\author{Hong-Rong Li$^{1}$}
\author{Franco Nori$^{2,4}$}
\affiliation{$^1$ Institute of Quantum Optics and Quantum
Information, School of Science,
Xi'an Jiaotong University, Xi'an 710049, China\\
$^2$ CEMS, RIKEN, Wako-shi, Saitama 351-0198, Japan\\
$^3$ Faculty of Physics, Adam Mickiewicz University, 61-614 Pozna\'n, Poland\\
$^4$ Physics Department, The University of Michigan, Ann Arbor,
Michigan 48109-1040, USA}
\date{\today}

\begin{abstract}
Phonon blockade is a purely quantum phenomenon, analogous to
Coulomb and photon blockades, in which a single phonon in an
anharmonic mechanical resonator can impede the excitation of a
second phonon. We propose an experimental method to realize phonon
blockade in a driven harmonic nanomechanical resonator coupled to
a qubit, where the coupling is proportional to the second-order
nonlinear susceptibility $\chi^{(2)}$. This is in contrast to the
standard realizations of phonon and photon blockade effects in
Kerr-type $\chi^{(3)}$ nonlinear systems. The nonlinear coupling
strength can be adjusted conveniently by changing the coherent
drive field. As an example, we apply this model to predict and
describe phonon blockade in a nanomechanical resonator coupled to
a Cooper-pair box (i.e., a charge qubit) with a linear
longitudinal coupling. By obtaining the solutions of the steady
state for this composite system, we give the conditions for
observing strong antibunching and sub-Poissonian phonon-number
statistics in this induced second-order nonlinear system. Besides
using the qubit to produce phonon blockade states, the qubit
itself can also be employed to detect blockade effects by
measuring its states. Numerical simulations indicate that the
robustness of the phonon blockade, and the sensitivity of
detecting it, will benefit from this strong induced nonlinear
coupling.
\end{abstract}

 \maketitle

\section{Introduction}
Quantum mechanics enables many breakthroughs that classical
physics cannot reach. However, due to decoherence, there exist
huge obstacles when quantum theory is applied to macroscopic
systems. In recent years, nanomechanical fabricating technologies
have achieved tremendous progress and provide ideal platforms to
explore fundamental questions in quantum mechanics. Many efforts
\cite{LaHaye04,Blencowe04,Schwab05,Poot12,Aspelmeyer14} have been
made to approach the quantum \allowbreak limits of nanomechanical
resonators (NAMRs), for instance, ground-state
cooling~\cite{Aspelmeyer14,Teufel11} and preparing nonclassical
states (examples including superposition
states~\cite{Yin13,Tan13}, squeezed
states~\cite{Xue07,Seok12,Pirkkalainen15}, etc).

In experimental implementations, the fundamental frequencies of
NAMRs range from tens of Hz to several GHz, and as a result, the
thermal environment significantly affects the coherences of
mechanical modes. For NAMRs at microwave
frequencies~\cite{Huang03}, the quantum limit can be approached
via cryogenic means (in the range of $\sim $mK). If the energies
of mechanical quanta (phonons), are larger than their thermal
energy, the quantum behavior of mechanical modes might be
observed. However, for mechanical modes of much lower frequencies,
the quantum coherences are fragile to thermal environments.
Usually, NAMRs need to be cooled to reach their ground states via
methods such as side-band
cooling~\cite{Marquardt07,Xue-07,Grajcar08} or active feedback
cooling~\cite{Poggio07,ZhangJ09}. To operate the mechanical
motions effectively, NAMRs are often combined with other systems
to form hybrid systems. Examples include optomechanical
systems~\cite{Aspelmeyer14,Kippenberg08} and quantum
electromechanical systems \cite{Blencowe04,Xiang13}. In the
quantum regime, NAMRs can be employed in fields such as quantum
information processing~\cite{Xiong15} and quantum
metrology~\cite{Woolley08,Arash15}.

\subsection{Obstacles for observing robust phonon blockade}

Phonon blockade (PB)~\cite{Liu10,Didier11,Miranowicz16} is another
purely quantum phenomenon, in which a single phonon in a resonator
can impede the transmission of a second one. Phonon blockade is an
analog of another quantum phenomenon named photon blockade (see
Refs.~\cite{Imamoglu97,Miranowicz13,Liu14,Xu14,Zhou15,Liu16} and
references therein). The interest in photon blockade is also
motivated for realizing single-photon sources for
quantum-information processing. Compared with classical states,
phonon and photon blockade states are described by the
sub-Poissonian distribution and can be interpreted as nonlinear
quantum scissors~\cite{Miranowicz13,Miranowicz01}. Photon blockade
has theoretically been predicted in various systems and was first
observed in an optical cavity coupled to a single trapped
atom~\cite{Birnbaum05}. However, realizing and observing PB are
still challenging for the following reasons: (1) The key point for
conventional phonon (photon) blockade is the realization of a
large nonlinearity with respect to the decay rate of the system.
However, moving into the strong nonlinear regime often requires
strict conditions which are hard to realize in most systems. (2)
The quantum coherence for NAMRs is very easy to be destroyed by
any noisy thermal environment. (3) Detecting PB directly is also
challenging: The position displacement for NAMRs is too tiny to be
detected effectively under current experimental
techniques~\cite{Liu10,Didier11,Lambert08}. Moreover, the
zero-point fluctuations for massive objects will limit the
measurement accuracy.

To date, most of the studies on phonon and photon blockade are
mainly based on nonlinearity (Kerr-type third-order $\chi ^{(3)}$
nonlinearity~\cite{Liu10,Imamoglu97,Miranowicz13,Miranowicz16} and
second-order $\chi ^{(2)}$
nonlinearity~\cite{Zhou15,Majumdar13,Shen14}) and quantum
interference effects~\cite{Zhou15,Tang15}. Phonon and photon
blockade in $\chi ^{(2)}$ and $\chi ^{(3)}$ nonlinear systems both
originate from the energy-level shift of multi-excitation states.
To observe the blockade of the second excitation, the decoherence
rate should be much smaller than the nonlinear strength. Phonon
and photon blockade can also be generated by utilizing optimal
conditions for quantum interference: transition paths for the
multi-excitation states are destructive and will cancel each
other, leading to a small population of the second excitation.
Recently, a new mechanism called unconventional photon blockade
has been predicted~\cite{Leonski05,Liew10,Flayac13,Gerace14,Lemonde14}, in which
strong two-qubit entanglement and strong photon antibunching can be observed via the
destructive quantum interference effect even in the weak nonlinear
regime. Reference~\cite{Tang15} theoretically proposes how to
realize photon blockade via quantum interference effects in a
quantum-dot cavity (without any nonlinearity). However, to avoid
multiexcited states, the strengths and relative phases of the
drive fields must be perfectly fixed when employing quantum
interference to prepare blockade states.

\subsection{Summary of our proposal}
It should be noted that the nonlinearity can be intrinsic or
induced via ancillary systems
\cite{Liu10,Johansson14,Miranowicz16}. Phonon blockade in a
Kerr-type nonlinear system has been demonstrated in
Refs.~\cite{Liu10,Didier11,Miranowicz16}. Inspired by these works,
our goal in this paper is to study PB via an effective
second-order nonlinear coupling, which remains unexplored. To
obtain a second-order nonlinear coupling, a NAMR is assumed to be
coupled to a Cooper-pair box (i.e., a charge qubit)
\cite{Xiang13,Makhlin01,You11} with a linear longitudinal coupling
($\sigma _{z}$ coupling) \cite{Wang07,Kerman13,Didier}. As
discussed in Refs. \cite{Wilson10,Liu142,Zhao15,Garziano15},
coupling a resonator with a superconducting qubit of longitudinal
form will induce multiphonon (multiphoton) processes.

It is worth stressing that our system is based on the longitudinal
coupling, instead of the transverse coupling assumed in the Rabi
model under the rotating-wave approximation. To our knowledge, all
former papers on phonon blockade have been based on the
Jaynes-Cummings model in the dispersive limit (i.e., assuming
large detuning). Our proposal, in which a strong quadratic
coupling between the charge qubit and the NAMR can be induced by
choosing appropriate driving parameters of the charge qubit, has
the following advantages:

(1) The intrinsic and induced effective $\chi ^{(3)}$
nonlinearities are usually very weak (about three orders lower of magnitude
than the Jaynes-Cummings coupling strength)
\cite{Zhang15,L15,Liu14,Miranowicz16}. However, in this
longitudinal-coupling system, the second-order nonlinear strength
can be much stronger and, as a result, we can observe robust PB
using lower-order nonlinear interactions.

(2) Besides using the qubit to produce PB states, the qubit itself
can also be employed to detect PB.

(3) Comparing with the direct nonlinear coupling between a qubit
and a NAMR \cite{Zhou06}, this type of effective coupling does not
require the qubit and the NAMR to be resonant, and the effective
coupling strength can be adjusted by controlling the strength of
the drive field or the longitudinal coupling strength.

In this paper, we first describe the model Hamiltonian in Sec.~II.
Then, in Sec.~III, we obtain the theoretical results for steady
states and give conditions to observe strong sub-Poissonian phonon
statistics and strong phonon antibunching in our proposal. In
Sec.~IV, we demonstrate our numerical results and discuss how to
increase the robustness against different types of noise in this
composite system. In Sec.~V, we propose a method to detect PB in
this setup. Section~VI presents the conclusions.

\section{Model}
\subsection{Hybrid system to generate phonon blockade}

\begin{figure}[tbph]
\centering
\includegraphics[width=7.0cm]{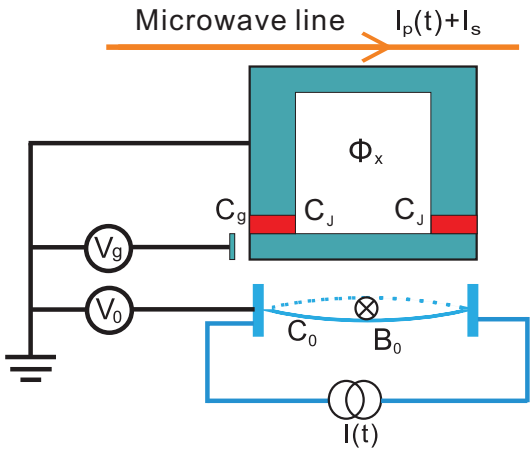}
\caption{(Color online) Schematic diagram of the coupled system of
a NAMR and a Cooper-pair box working as a charge qubit. The two
Josephson junctions (red rectangles) with tunneling energy $E_{J}$
and capacitance $C_{J}$ form a SQUID loop. The perpendicular
static magnetic field $B_{0}$, together with the current $I(t)$
through the NAMR, produces the Lorentz force to drive the NAMR.
Here a gate voltage $V_{g}\ $and a static voltage $V_{0}$ are
applied to the capacitances $C_{g}$ and $C_{0}$, respectively. The
microwave line is located right above the charge qubit. The
time-dependent current $I_{\text{p}}(t)$ and static current
$I_{\text{s}} $ in the microwave line induce a magnetic flux $\Phi
_{x}$ through the SQUID loop.} \label{fig1}
\end{figure}

The proposed setup is illustrated in Fig.~\ref{fig1}. A lossless
NAMR (with a fundamental frequency $\omega _{0}$, an effective
length $L$, and a mass $m $) is coupled to a charge qubit by
applying a static voltage $V_{0}$ through a capacitance $C_{0}(x)$
\cite{Armour02,Irish03,Sun06}. Around its equilibrium position,
$x=0$, $C_{0}(x)$ depends on the displacement $x$ of the NAMR, and
the distance between the charge qubit and NAMR is $d$. The
tunneling energy and capacitance of two Josephson junctions in a
superconducting quantum interference device (SQUID) are $E_{J}$
and $C_{J}$, respectively. The charge states of the qubit can be
precisely tuned by adjusting the gate voltage $V_{g}$ and the
static voltage $V_{0}$. The driving force for the NAMR is induced
by a time-dependent current $I(t)=I_{0}\cos (\omega _{f}t)$ and a
perpendicular static magnetic field $B_{0}$~\cite{Liu10}. The
magnetic flux $\Phi _{x}$ through the SQUID loop is produced by
the microwave field in the microwave line. Thus, we express the
Hamiltonian of the total system as
\begin{eqnarray}
H_{\text{total}} &=&H_{\text{Q}}+H_{\text{NAMR}},\label{eq1} \\
H_{\text{Q}} &=&2E_{c}\left( 2n_{g}-1\right) \sigma
_{z}-E_{J}\cos\!\left(
\frac{\pi \Phi _{x}}{\Phi _{0}}\right) \sigma _{x}, \\
H_{\text{NAMR}} &=&\hbar \omega _{0}b^{\dag }b+ \hbar \epsilon \left( b^{\dag }+b\right)\cos\left(\omega
_{f}t\right),
\end{eqnarray}
where $H_{\text{Q}}$ is the Hamiltonian of the charge qubit. In
the neighborhood of the charge-degeneracy point
$n_{g}=[C_{g}V_{g}+C_{0}(0)V_{0}]/\left( 2e\right) \simeq 0.5$,
the dephasing noise on the qubit will be suppressed. The charge
qubit can be characterized by the pseudospin Pauli operators
$\sigma _{z}=|e\rangle \langle e|-|g\rangle \langle g|$ and
$\sigma _{x}=\sigma _{+}+\sigma _{-}=|e\rangle \langle
g|+|g\rangle \langle e|,$ with $|e\rangle $ ($|g\rangle $) being
the excited (ground) state for the qubit. Here $H_{\text{NAMR}}$
is the Hamiltonian for the driven mechanical mode of the NAMR, and
$b^{\dag }$ and $b$ are the phonon creation and annihilation
operators, respectively. We define $\hbar \omega
_{\text{q}}$=$4E_{c}\left( 2n_{g}-1\right) $, with
$E_{c}=e^{2}/\left( 2C_{\Sigma }\right) $ being the charging
energy of the qubit with the total capacitance $C_{\Sigma
}=C_{J}+C_{0}(0)+C_{g}$ (we assume that $C_{0}(0)$ and $C_{g}$ are
much less than $C_{J}).$ The displacement $x$ of the NAMR gives
rise to the linear modulation of the capacitance between the NAMR
and Cooper-pair box island, that is, $C_{0}(x)=C_{0}(0)\left(
1-x/d\right).$ This leads to the coupling constant
\begin{equation}
g=\frac{2E_{c}C_{0}(0)V_{0}}{ed\hbar }X_{0}\text{,}  \label{eq2}
\end{equation}
where $X_{0}=\sqrt{\hbar /\left( 2m\omega _{0}\right) }$ describes
the zero-point fluctuations of the NAMR. The parameter
\begin{equation}
\epsilon =\hbar^{-1}B_{0}I_{0}LX_{0}
\end{equation}%
is the driving strength of the Lorentz force for the NAMR, which
is induced by an alternating current at a frequency $\omega _{f}$
and a static magnetic field $B_{0}$.

The second term in $H_{\text{Q}}$ is the Josephson energy. Here
$\Phi _{0}=\hbar /\left( 2e\right) $ is the flux quantum. The
coupling between the microwave line and the qubit results from the
magnetic flux
\begin{equation}
\Phi_{x}=M[I_{\text{p}}(t)+I_{\text{s}}]
\end{equation}
applied through the SQUID loop via the mutual inductance $M.$ Here
$I_{\text{p}}(t)=\varepsilon _{\text{p}}\cos (\omega
_{\text{p}}t)$ and $I_{\text{s}}$ are the time-dependent and
direct current in the microwave line, respectively. Under the
conditions $MI_{\text{s}}=\Phi _{0}/2$ and $I_{\text{p}}(t)\ll
I_{\text{s}}$, we expand the Josephson energy to first order as
$E_{J}\sigma _{x}\sin y \approx E_{J}y \sigma _{x}$, where $y=\pi
MI_{\text{p}}(t)/\Phi _{0}\ll 1$. Applying a frame rotating at a
frequency $\omega _{\text{p}}$ and adopting the rotating wave
approximation, the Hamiltonian of the total system becomes
(setting $\hbar =1$)
\begin{equation}
\begin{split}
H_{\text{total}} &=\frac{1}{2}\Delta \sigma _{z}+\omega
_{0}b^{\dag
}b+g\sigma _{z}(b^{\dag }+b) \\
&\quad+\Omega _{\text{p}}\left( \sigma _{+}+\sigma _{-}\right)
+\epsilon \left( b^{\dag }e^{-i\omega _{f}t}+be^{i\omega
_{f}t}\right) ,
\end{split}
\label{eq3}
\end{equation}%
where $\Delta =\omega _{\text{q}}-\omega _{\text{p}}$ is the
drive-excitation detuning and
\begin{equation}
\Omega _{\text{p}}=\frac{\pi E_{J}MI_{\text{p}}(t)}{\hbar \Phi
_{0}}
\end{equation}%
is the effective Rabi frequency of the drive field.

\subsection{Induced nonlinear qubit-NAMR coupling}

To achieve the nonlinear coupling, we set the driving field for
the qubit as red sideband with $\Delta \simeq 2\omega _{0}$, as
shown in Fig.~\ref{fig2}(a) (black arrows). Moreover, we do not
consider the weak drive $\epsilon $ of the NAMR at the beginning.
Performing the polariton
transformation~\cite{Liu142,Wilson04,Wilson} to $H_{\text{total}}$
\begin{equation}
\tilde{H}=e^{S}H_{\text{total}\;}e^{-S}
\end{equation}%
with $S=\beta \sigma _{z}(b^{\dag }-b)$ and $\beta =g/\omega
_{0}$, we obtain
\begin{equation}
\tilde{H}=\frac{1}{2}\Delta \sigma _{z}+\omega _{0}b^{\dag
}b+[\Omega _{\text{p}}\sigma _{+}e^{2\beta (b^{\dag
}-b)}+\text{H.c.}],  \label{eqc1}
\end{equation}%
where H.c. denotes the Hermitian conjugate of the last term in
$\tilde{H}$. In this shifted oscillator framework, we expand the
above equation to second order in the small parameter $\beta $,
and we obtain
\begin{equation}
\begin{split}
\tilde{H} &=\frac{1}{2}\Delta \sigma _{z}+\Omega
_{\text{p}}(\sigma
_{+}+\sigma _{-})+\omega _{0}b^{\dag }b \\
&\quad +2\Omega _{\text{p}}[\beta \sigma _{+}(b^{\dag }-b)+\text{H.c.}] \\
&\quad +2\Omega _{\text{p}}[\beta ^{2}\sigma _{+}(b^{\dag
}-b)^{2}+\text{H.c.}].
\end{split}
\label{eqc2}
\end{equation}
The second term in Eq.~(\ref{eqc2}) will cause the dynamical
energy shifts for the charge qubit. Defining the shifted energy as
\begin{equation}
\tilde{\Delta}=\sqrt{\Delta ^{2}/4+\Omega _{\text{p}}^{2}},
\label{TildeDelta}
\end{equation}
and expressing the new eigenstates in the basis of the charge
states
\begin{eqnarray}
\begin{split}
|-\rangle &=\cos \frac{\theta }{2}|g\rangle -\sin \frac{\theta }{2}%
|e\rangle ,   \\
|+\rangle &=\sin \frac{\theta }{2}|g\rangle +\cos \frac{\theta }{2}%
|e\rangle
\end{split}
\label{PlusMinusStates}
\end{eqnarray}%
with $\tan \theta =2\Omega _{\text{p}}/\Delta $, we rewrite
$\tilde{H}$ as

\begin{equation}
\setlength{\abovedisplayskip}{2pt}
\begin{split}
\tilde{H} &=\tilde{\Delta}\tilde{\sigma}_{z}+\omega _{0}b^{\dag
}b+2\lambda
\Omega _{\text{p}}(\tilde{\sigma}_{+}-\tilde{\sigma}_{-})(b^{\dag }-b) \\
&\quad +2\Omega _{\text{p}}\lambda ^{2}\sin \theta
\tilde{\sigma}_{z}(b^{\dag
}-b)^{2} \\
&\quad +2\Omega _{\text{p}}\beta ^{2}\cos \theta (\tilde{\sigma}_{+}+\tilde{\sigma}%
_{-})(b^{\dag }-b)^{2},
\end{split}
\label{eqc3}
\end{equation}%
where $\tilde{\sigma}_{z}=|+\rangle \langle +|-|-\rangle \langle
-|$, $\tilde{\sigma}_{+}=|+\rangle \langle -|\ $and
$\tilde{\sigma}_{-}=|-\rangle \langle +|$. The driving for the
qubit is far detuned with $\Delta \gg \Omega _{\text{p}}$, so the
rotating angle $\theta $ of the new basis is very small, and we
have $\sin \theta \simeq \Omega _{\text{p}}/\omega _{0}\ll 1$ and
$\cos \theta \simeq 1$. The third and forth terms in
Eq.~(\ref{eqc3}) represent the energy shifts for the qubit and the
NAMR. Assuming that the qubit is approximately in its ground state
($\sigma _{z}=\tilde{\sigma}_{z}=-1,$ since the sideband driving
is far detuned), the renormalized NAMR frequency can be expressed
by
\begin{equation}
\quad \omega _{0}^{{\prime }} =\omega _{0}-\frac{4\Omega
_{\text{p}}^{2}\lambda ^{2}} {3\omega _{0}^{3}},
\label{RenormalizedFrequency}
\end{equation}
where the second term in this equation describes the
eigenfrequency shift of the NAMR~\cite{Irish03}. In this paper, we consider the resonant case
$\tilde{\Delta}=\omega _{0}^{{\prime }}$ and assume the parameters
satisfy the condition
\begin{equation}
\Delta \simeq 2\omega _{0}\ \gg\ \max \{g,\Omega _{\text{p}}\}\
\ge\ \min \{g,\Omega _{\text{p}}\}\ \gg \epsilon ,
\label{Condition1}
\end{equation}%
Performing the unitary transformation
\begin{equation}
U=\exp [-i(\tilde{\Delta}\tilde{\sigma}_{z}+\omega _{0}^{{\prime
}}b^{\dag }b)t]
\end{equation}
of $\tilde{H}$ in Eq.~(\ref{eqc3}), we can neglect the rapid
oscillating terms and obtain the effective Hamiltonian as follows

\begin{equation}
\tilde{H}=\lambda(\tilde{\sigma}_{+}b^{2}+\tilde{\sigma%
}_{-}b^{\dag 2}).  \label{eqc4}
\end{equation}
where
\begin{equation}
\lambda =\frac{2\Omega _{\text{p}}g^{2}}{\omega _{0}^{2}}
\label{eqcouple}
\end{equation}%
is the effective nonlinear coupling strength between the states
$|g,n+2\rangle $ and $|e,n\rangle$, where the $|n\rangle $ are the
Fock states of the mechanical mode. Equation.~(\ref{eqc4})
describes a nonlinear process that the qubit can be excited by the
annihilation of two phonons, or the inverse process as shown in
Fig.~\ref{fig2}(b).

It should be noted that $\tilde{H}$ in Eq.~(\ref{eqc4}) is the
Hamiltonian after performing the small polariton transformation
and rotating the qubit basis with an angle $\theta,$ so the
evolution of the density matrix $\tilde{\rho}(t),$ described by
$\tilde{H},$ is also in the rotating frame. In fact, we are
ultimately interested in the dynamics as seen in the original
non-transformed lab frame $\rho (t).$ To obtain a complete
description of the system, we also need to apply these two
rotating transformations to the bath-system coupling. In
principle, this can be done by transforming the operators in the
master equation. Only after these two steps we can obtain the
explicit solutions to this system. However, here we have assumed
that both of these two transformations just \emph{slightly} rotate
the density matrix under the small parameters $\beta $ and $\theta
$, so $\tilde{\rho}(t)$ can be viewed as an approximate solution
for $\rho (t).$ For the collapse operators in the master equation,
we can also neglect the terms of order $\lambda $ and $\theta $.
We find that these assumptions to be valid by comparing our
theoretical and numerical results.

By including the drive of the NAMR, the effective
Hamiltonian of the system can be expressed as
\begin{equation}
H_{\text{eff}}=\lambda (b^{2}\sigma _{+}+b^{\dag 2}\sigma
_{-})+\epsilon (b^{\dag }e^{i\Delta _{d}t}+be^{-i\Delta _{d}t}),
\label{eq4}
\end{equation}%
where $\Delta _{d}=\omega _{0}^{{\prime }}-\omega _{f}$ is the
detuning between the renormalized NAMR frequency, given by
Eq.~(\ref{RenormalizedFrequency}), and the frequency $\omega _{f}$
of the alternating current $I(t)$. Assuming $g=\Omega
_{\text{p}}=0.1\omega _{0}$, a strong nonlinear coupling with
strength $\lambda =\omega _{0}/500$ can be achieved. Defining the
time-dependent probabilities $P_{ij}$ ($i=0,1,2...$ and $j=g,e)$
for the states $|j,i\rangle $ and setting $\epsilon =0$, we
numerically solve the Schr\"{o}dinger equation governed by
$H_{\text{total}}$ with the initial state $|e,0\rangle ,$ and plot
the results in Fig.~\ref{fig2}(d). We find that the amplitudes of
$P_{0e}$ and $P_{2g}$ approximately exhibit Rabi oscillations with
the Rabi frequency $\sqrt{2}\lambda .$

Thus, under the appropriate red-sideband driving for the qubit, we
realize an effective nonlinear coupling between the NAMR and the
qubit. Different from the direct nonlinear coupling proposal in
Ref.~\cite{Zhou06}, the induced nonlinear coupling described here
should be easier to realize in experiments. In this work, we
consider a pseudospin charge qubit coupling with a NAMR of the
longitudinal form just as an example, but the same coupling also
exists between mechanical modes and a semiconductor quantum
dot~\cite{Wilson04} (or a carbon nanotube~\cite{Wilson}, an
electronic-spin qubit~\cite{Rabl09}, etc.). Thus, our discussions
in this paper can also be applied to those systems.

\section{Analytical description of phonon blockade}

\begin{widetext}
\begin{figure*}[tbph]
\centering
\includegraphics[width=0.97\linewidth]{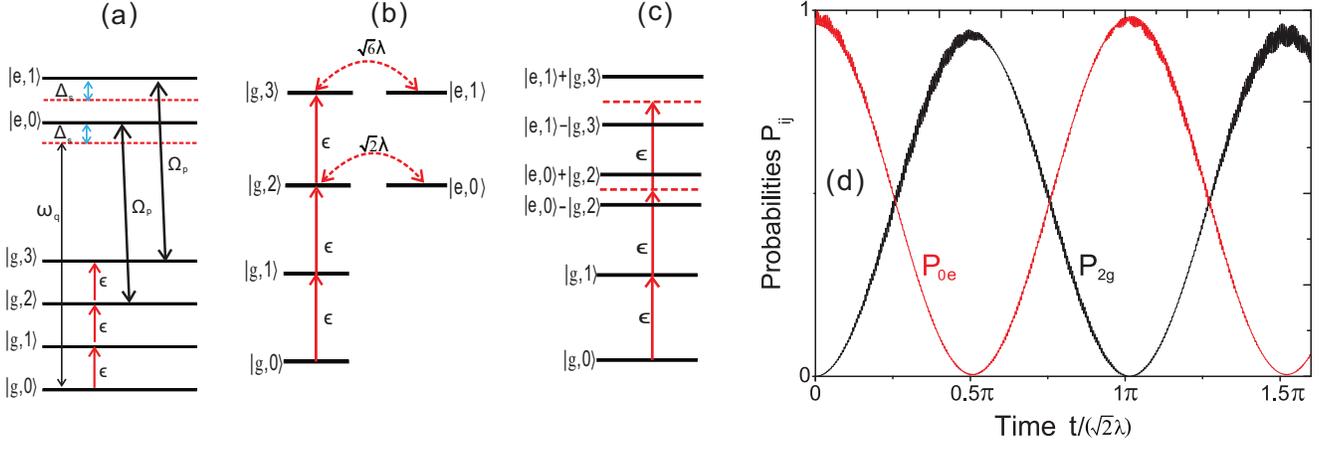}
\caption{(Color online) {Schematic energy-level diagrams [in (a),
(b), and (c)]} explaining the occurrence of phonon blockade in our
approach. The red and black arrows represent the drivings for the
NAMR and qubit with strengths $\protect\epsilon $ and $\Omega
_{\text{p}}$, respectively. (a) The energy shift $\Delta
_{s}=\tilde{\Delta}-\Delta $ (blue two-direction arrows) of the
energy gap for the qubit is due to the side-band driving. (b) The
induced nonlinear strength $\protect\lambda $ leads to the
transfer between states $|e,0\rangle \leftrightarrow |g,2\rangle $
($|e,1\rangle \leftrightarrow |g,3\rangle $) with rate
$\protect\sqrt{2}\protect\lambda $
($\protect\sqrt{6}\protect\lambda $). (c) Due to the energy level
splitting between the dressed states ($|e,n\rangle \pm
|g,n+2\rangle)/\protect\sqrt{2}$, there is no corresponding
transition level for a multiphonon being excited. (d) The time
evolution of the probabilities $P_{2g}$ (black curve) and $P_{0e}$
(red curve) demonstrate the Rabi oscillations between the states
$|g,2\rangle $ and $|e,0\rangle $ without considering any decay
channels.} \label{fig2}
\end{figure*}
\end{widetext}

Once the effective nonlinear coupling in Eq.~(\ref{eq4}) is
induced, the ground state is $|g,0\rangle $ and the first-excited
state is $|g,1\rangle $. However, the second excited states are
the superposition states of $|g,2\rangle $ and $|e,0\rangle $ with
splitting $2\sqrt{2}\lambda .$ The energy-level diagrams for the
first few excitation states can be found analytically as shown in
Fig.~\ref{fig2}(b, c). Hence, when the NAMR is driven by a
resonant force with strength $\epsilon $ (red solid line in
Fig.~\ref{fig2}), the first phonon of the NAMR can be easily
generated, while the second phonon can be hardly excited, since
there are no corresponding transmission energy levels for the
second incoming phonon. Thus, the second phonon is blocked by the
first incoming phonon.

However, for a strong driving rate of the NAMR, high excitation
states might become occupied and the blockade effect will be
weaker. Moreover, the decoherences of the qubit and the NAMR also
lead to the deterioration of PB. In this section, we will derive
the expression of the second-order correlation function $g_{2}(0)$
under the low-excitation approximation, and give the conditions
for observing strong antibunching and sub-Poissonian phonon
statistics effects.

First we analyze the steady state of our system by analyzing
non-Hermitian Hamiltonians.

\subsection{Steady-state solutions of the Schr\"odinger equation for the non-Hermitian Hamiltonian}

If we assume that the driving strength $\epsilon $ is much smaller
than the effective nonlinear coupling strength $\lambda ,$
then we can conjecture that the infinite-dimensional
mechanical-mode Hilbert space can be reduced into the
two-phonon excitation subspace. Then we can expand the wave
function in the basis spanned by $\left\{|g,0\rangle, |g,1\rangle,
|e,0\rangle, |g,2\rangle \right\}$ as follows:
\begin{equation}
|\psi \rangle =C_{0g}|g,0\rangle +C_{1g}|g,1\rangle
+C_{0e}|e,0\rangle +C_{2g}|g,2\rangle ,  \label{eq5}
\end{equation}%
with $|C_{ij}|^{2}=P_{ij},$ which is a good approximation to
describe phonon blockade. The validity of this approximation
will be discussed in Sec.~IV, e.g., in the analysis of
Fig.~\ref{fig8}. With the assumption $\Delta _{d}=0$, the decay of the charge qubit and the mode loss
can be treated by the non-Hermitian time-independent Hamiltonian
\begin{equation}
H_{\text{nH}}=H_{\text{eff}}-i\frac{\kappa }{2}b^{\dag
}b-i\frac{\Gamma }{2} |e\rangle \langle e|,  \label{eq6}
\end{equation}%
where we have assumed that the qubit and the NAMR couple with the
vacuum reservoirs, and the corresponding damping rates are $\Gamma
$ and $\kappa $, respectively. Substituting the state $|\psi
\rangle $ and $H_{\text{nH}}$ to the Schr\"{o}dinger equation, we
obtain the equations of motion for the coefficients in
Eq.~(\ref{eq5}):
\begin{equation}
\begin{split}
i\frac{\partial }{\partial t}C_{1g} &=\epsilon C_{0g}-i\frac{\kappa }{2}%
C_{1g}+\sqrt{2}\epsilon C_{2g}, \\
i\frac{\partial }{\partial t}C_{2g} &=\sqrt{2}\epsilon C_{1g}-i\kappa C_{2g}+%
\sqrt{2}\lambda C_{0e}, \\
i\frac{\partial }{\partial t}C_{0e} &=\sqrt{2}\lambda C_{2g}-i\frac{\Gamma }{%
2}C_{0e}.
\end{split}
\label{eq7}
\end{equation}

By setting $\partial C_{ij}/\partial t=0$ ($i=$0, 1, 2 and
$j$=$g$, $e$), we obtain the steady-state solution for each
coefficient:
\begin{subequations}
\begin{align}
C_{0e}& =\frac{2\sqrt{2}\lambda }{i\Gamma }C_{2g},  \label{eq8a} \\
C_{2g}& =\frac{\sqrt{2}\epsilon }{i\left( 2\kappa +\frac{4\lambda ^{2}}{%
\Gamma }\right) }C_{1g},  \label{eq8b} \\
C_{0g}& =\frac{i\kappa }{2\epsilon }C_{1g}-\sqrt{2}\epsilon
C_{2g}. \label{eq8c}
\end{align}
For a strong phonon blockade effect, the NAMR is treated as a
two-level system while multiphonon states ($n\geq 2)$ are
suppressed by the qubit-NAMR nonlinear coupling; so we assume that
\end{subequations}
\begin{equation*}
|C_{2g}|^{2}\ll \min \{|C_{0g}|^{2},|C_{1g}|^{2}\}
\end{equation*}
and $|C_{0g}|^{2}+|C_{1g}|^{2}\simeq 1$. Neglecting $C_{2g}$ in
Eq.~(\ref{eq8c}), we obtain
\begin{equation}
|C_{1g}|^{2}=\frac{4\epsilon ^{2}}{8\epsilon ^{2}+\kappa ^{2}},
\label{eq9}
\end{equation}%
and according Eq.~(\ref{eq8b}), we have
\begin{equation}
|C_{2g}|^{2}=\frac{8\epsilon ^{4}}{\left( 2\kappa +4\lambda
^{2}/\Gamma \right) ^{2}\left( 8\epsilon ^{2}+\kappa ^{2}\right)
}.  \label{eq10}
\end{equation}%

These results will be useful to calculate second-order correlation
function in the following sections.
\subsection{Sub-Poissonian phonon statistics}

The sub-Poissonian phonon statistics can be revealed by measuring
the zero-delay-time second-order correlation function
\begin{equation}
g_{2}(t,0) =\frac{\langle b^{\dag }(t)b^{\dag }(t)b(t)b(t)\rangle
}{\langle b^{\dag }(t)b(t)\rangle^2}. \label{eq11a}
\end{equation}%
We recall that if a given phonon state is described by
$g_{2}(t,0)<1$ [$g_{2}(t,0)>1$], then it exhibits sub-Poissonian
(super-Poissonian) phonon statistics, which is also sometimes
referred to as phonon antibunching (bunching).

However, here we refer to phonon antibunching and bunching in a
more common way, as defined in Sec. IV.B.

Let us denote $g_{2}(0)=\lim_{t\rightarrow
\infty}g_{2}(t,0)$. Then for the state, given by Eq.~(\ref{eq5}), we find that the
correlation function $g_{2}(0)$ can be expressed by the
probability amplitudes as follows
\begin{equation}
g_{2}(0)\simeq \frac{2|C_{2g}|^{2}}{|C_{1g}|^{4}}=\frac{8\epsilon
^{2}+\kappa ^{2}}{\left( 2\kappa +4\lambda ^{2}\Gamma^{-1} \right)
^{2}}. \label{eq12}
\end{equation}

From Eq.~\!\!(\ref{eq12}), we conclude that, in our second-order
nonlinear system, the effective nonlinear coupling strength
$\lambda $ and the qubit decay rate $\Gamma $ significantly affect
the dip of the sub-Poissonian phonon statistics. If a relatively
strong coupling strength $\lambda $ is induced and the relation
\begin{equation}
\frac{4\lambda ^{2}}{\Gamma} \gg \max \{2\sqrt{2}\epsilon ,\kappa \}
\label{StrongCouplingCondition}
\end{equation}
is satisfied, strong sub-Poissonian phonon statistics can be
observed in this system with $g_{2}(0)\ll 1.$
\section{Numerical description of phonon blockade}

\subsection{Steady-state solutions of the master equation}

In this section we numerically study the phonon blockade effect
via the standard master equation approach. Numerical computations
were performed using the Python package QuTiP~\cite{Johansson13}.
We perform numerical calculations in the Fock space of the
NAMR of dimension $M$=10, which is much larger than that
assumed in Eq.~(\ref{eq5}). To verify our analytical results of
phonon blockade in Sec.~III, we adopt the original Hamiltonian
$H_{\text{total}}$ in Eq.~(\ref{eq3}) (not $H_{\text{eff}}$) to
proceed with our numerical simulations. The
Kossakowski-Lindblad master equation for the reduced density
matrix $\hat{\rho}(t)$ of the system reads
\begin{eqnarray}
\frac{d\hat{\rho}(t)}{dt}
&=&-i[H_{\text{total}},\hat{\rho}(t)]+D[\sigma
_{-},\Gamma ]\hat{\rho}(t)\notag \\
&&+\,D[\sigma _{z},\Gamma
_{f}/2]\hat{\rho}(t)+n_{\text{th}}D[b^{\dag },\kappa
]\hat{\rho}(t)  \notag
\\
&&+\,(n_{\text{th}}+1)D[b,\kappa ]\hat{\rho}(t), \label{eq13}
\end{eqnarray}%
where
\begin{equation}
D[A,\Omega ]\hat{\rho}=\frac{1}{2}\Omega (2A\hat{\rho}A^{\dag }-A^{\dag }A%
\hat{\rho}-\hat{\rho}A^{\dag }A)
\end{equation}%
are Lindblad-form terms, and $\Gamma _{f}$ is the pure dephasing
rate of the qubit. Recall that $\Gamma $ corresponds to the qubit
decay rate and $\kappa$ the decay rate of the NAMR. In this
proposal, the NAMR might couple to a thermal reservoir of
temperature $T$ with thermal phonon number
$n_{\text{th}}=\{\exp[\hbar \omega _{0}/(k_{B}T)]-1\}^{-1}$, where
$k_{B}$ is the Boltzmann constant.
\subsection{Quantum signatures for steady-state phonon blockade: Phonon antibunching
and sub-Poissonian phonon statistics}

\begin{figure}[tbph]
\centering \includegraphics[width=8.5cm]{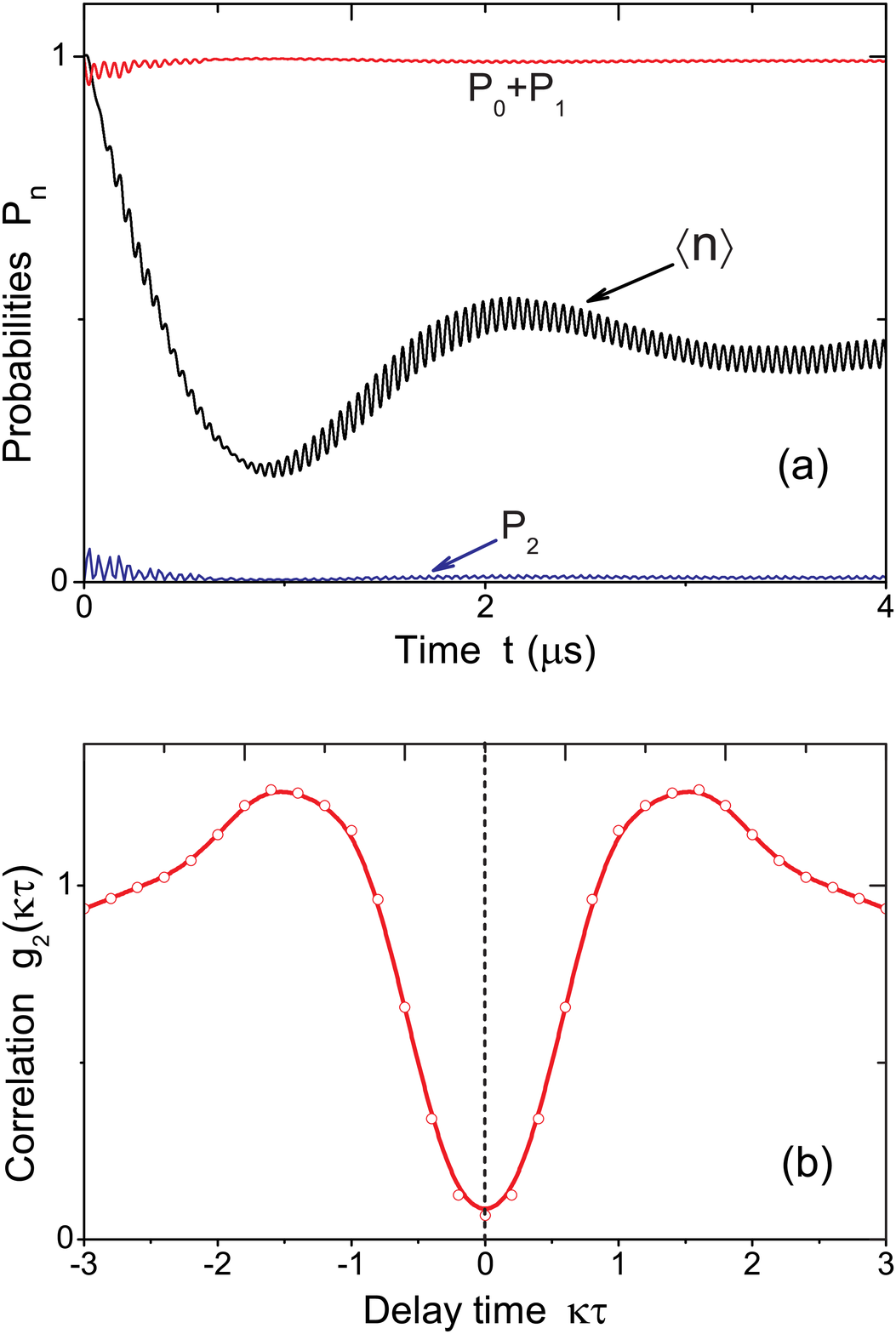} \caption{(Color
online) (a) Probabilities $P_{n}$ of measuring $n$ phonons and the
average phonon number $\langle n\rangle$ as a function of the
evolution time. (b) Phonon antibunching is revealed by the
second-order correlation function versus the rescaled delay time
$\protect\kappa \protect\tau .$ Other parameters are: $\Gamma
/(2\protect\pi )=1$~MHz, $\Gamma _{f}=0$, $\epsilon /(2\protect\pi
)=0.2$~MHz, $\Delta _{d}=0$, and $n_{\text{th}}=0.$} \label{fig3}
\end{figure}

Here we show that phonon blockade in the infinite-time limit of
the dissipative system is also a nonclassical effect because the
generated steady-state of the NAMR can exhibit both phonon
antibunching and sub-Poissonian phonon statistics.

To reveal phonon antibunching, we apply the steady-state
second-order correlation function
\begin{equation}
g_{2}(t,\tau)=\frac{\langle b^{\dag }(t)b^{\dag }(t+\tau )b(t+\tau
)b(t)\rangle }{\langle b^{\dag }(t)b(t)\rangle \langle b^{\dag
}(t+\tau )b(t+\tau )\rangle }, \label{eq11b}
\end{equation}%
where $\tau $ is the time delay between two measurements. This
function reduces to Eq.~(\ref{eq11a}) for $\tau$=0. The two-time
correlation function of the steady state of the mechanical mode is
the function of only the time delay as given by
\begin{equation}
g_{2}(\tau )=\lim_{t\rightarrow \infty}g_{2}(t,\tau). \label{eq26}
\end{equation}%

Here we refer to phonon antibunching (bunching) for a phonon
steady state only if $g_{2}(\tau )>g_{2}(0)$ [$g_{2}(\tau
)<g_{2}(0)$] for a delay time $0<\tau $, according to the standard
definition of these effects~\cite{MandelBook}. Phonon antibunching
also occurs if $g^{(2)}(\tau )$ has a strict local minimum at
$\tau =0$~\cite{Miran99}. Here we only study the phonon
antibunching of stationary states. Note that such effect can also
be observed for nonstationary cases, but a modification of this
definition would be required~\cite{Miran99}.

Note that these phonon antibunching and bunching effects are
defined via two-time phonon-number correlations, while the
sub-Poissonian and super-Poissonian statistics are given via
single-time phonon-number correlations. So, these are different
effects and a given state of the NAMR can be~\cite{Zou90} either
(1) both sub-Poissonian and antibunched, or (2) sub-Poissonian and
bunched, or (3) super-Poissonian and antibunched, or (4)
super-Poissonian and bunched. In this paper we focus on the case
(1). Finally, we stress that sub-Poissonian statistics and
antibunching are purely nonclassical effects, since they
correspond to the violation of classical
inequalities~\cite{Miran10}. Thus, the observation of either of
these effects can be a signature of the quantumness of a NAMR.

In current experiments with a charge qubit and a NAMR, their
coupling strength is of orders from tens to hundreds of MHz, as
shown in Refs.~\cite{MartinI04,RablP04,tianl05}. Moreover, the
quality factor of a NAMR at microwave frequencies is around
$10^{3}\sim10^{5}$~\cite{Teufel11,Sidles95}. The decay rate of a
charge qubit can reach the order of 1~MHz~\cite{Schuster07}. In
the following discussion, we assume that the NAMR oscillates
at $\omega _{0}/(2\pi )$=1~GHz with quality factor $Q$=5$\times
$10$^{3}$ [$\kappa /(2\pi )=0.2$~MHz]\negthinspace\,
and the coupling strength with the qubit is $g/(2\pi )=80$~MHz.
Under the driving rate $\Omega _{\text{p}}/(2\pi )=100$~MHz, the
effective nonlinear coupling strength is $\lambda /(2\pi
)=1.28$~MHz. Defining the time-dependent probabilities
$P_{n}(t)=\text{Tr}[|n\rangle \langle n|\hat{\rho}(t)]$ for the
phonon number $n$ and mean phonon number $\langle n\rangle
=\text{Tr}[b^{\dag }b\hat{\rho}(t)]$, we have numerically solved
the master equation and the results are shown in
Fig.~\ref{fig3}(a). We find that the sum of $P_{0}$ and $P_{1}$ is
almost one, while $P_{2}$ is of an extremely low amplitude,
indicating that phonon blockade occurs in this hybrid system.
Moreover, the average phonon number $\langle n\rangle $ of the
steady-state oscillates around $\sim $ \negthinspace 0.44, which
is due to the high-order terms, as shown in Sec.~II.B. In the
following numerical results, without loss of generality, we adopt
the ensemble-average $\langle n\rangle $ and $P_{2}$ of the steady
state to calculate $g_{2}(0).$

In addition to the time-dependent probabilities $P_{n}(t),$ other
quantum signatures for PB are the phonon intensity correlations
with finite-time delays. The delay-time-dependent second-order
correlation functions $g_{2}(\tau )$ are plotted in
Fig.~\ref{fig3}(b), from which it can be found that if the time
delay between two measurements $\kappa \tau \ $is within $\sim $1,
then a phonon antibunching dip can be observed.

\begin{figure}[tbph]
\centering
\includegraphics[width=8.75cm]{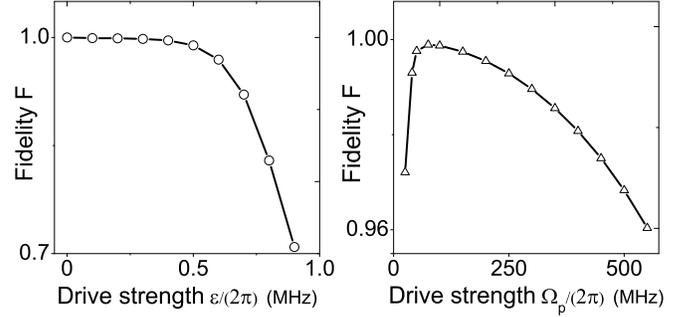}
\caption{The fidelity $F$, given by Eq.~(\ref{eqtrucation}),
of state truncation vs (a) the drive strength $\epsilon$ and (b)
the effective Rabi frequency $\Omega_{\text{p}}$ of the drive
field. The other parameters used here are the same as those in
Fig.~\protect\ref{fig3}(a).} \label{fig8}
\end{figure}

In Sec.~III.A, we assumed that the Hilbert space of the
system is truncated into its subspace spanned by the four states,
given in Eq.~(\ref{eq5}). This approximation will be not valid if
the higher-energy levels are excited. As shown in
Ref.~\cite{Miranowicz13}, we use the fidelity of state truncation
to estimate the quality of this effect in a phonon blockade system. With our
precise numerical solution obtained in a larger Hilbert space, the
fidelity is the sum of the steady-state probabilities of the states
considered, which is defined as
\begin{equation}
F(\rho_{ss})=|C_{0,g}|^{2}+|C_{1,g}|^{2}+|C_{2,g}|^{2}+|C_{0,e}|^{2},
\label{eqtrucation}
\end{equation}%
where $\rho_{ss}$ is the steady state of the system. The fidelity
$F\approx$\! 1 indicates that we can effectively use these four
states to expand the density matrix of the system, and the
analytical results in Sec.~III are valid. Otherwise, if the
fidelity $F$ is much smaller than 1, we can conclude that the
higher energy-level beyond these four states are effectively
occupied, and it is not enough to use only these four states to
describe the system.

In Fig.~\ref{fig8}(a), we plot the fidelity $F$ as a function
of the drive strength $\epsilon$ of the NAMR. With increasing
$\epsilon$, the fidelity $F$ does not change much (and remains
above 0.99) if $\epsilon/2\pi\! <$\! 0.5~MHz. However, if we
continue increasing the drive strength $\epsilon$, the fidelity
starts to decrease rapidly. In this case, the nonlinear effects
cannot effectively prevent multiphonon states being excited. As a
result, the phonon blockade effects will be destroyed.

The relation between the fidelity $F$ and the pump strength
$\Omega_{\text{p}}$ of the qubit is shown in Fig~\ref{fig8}(b). We
find that the fidelity initially increases rapidly with the pump strength
$\Omega_{\text{p}}$. This is because the nonlinear
strength goes up as shown in Eq.~(\ref{eqcouple}). Therefore,
blockade effects will be enhanced. However, when
$\Omega_{\text{p}}/(2\pi)$ increases more than 100~MHz, then the
fidelity starts to decrease. This is because the large detuning
approximation $\Delta \simeq 2\omega _{0}\! \gg \Omega
_{\text{p}}$, given by Eq.~(\ref{Condition1}), is not valid any
more when $\Omega_{\text{p}}$ is too strong. Consequently, the
qubit can be effectively excited and the transitions $|g,1\rangle
\leftrightarrow |e,1\rangle \leftrightarrow |g,3\rangle$ can
occur. Thus, multiphonon states are effectively excited. Moreover,
higher-order terms will also deteriorate the fidelity of
truncation, as shown in Sec.~II.B. It is seen that there is a
trade-off when we want to induce the strong nonlinear coupling
between the NAMR and the qubit while limiting the strength of the
qubit drive to a certain regime.

\begin{figure}[tbph]
\centering
\includegraphics[width=8.8cm]{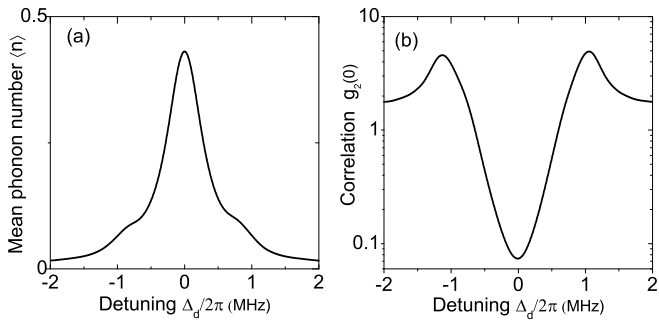}
\caption{(a) Mean phonon number $\langle n\rangle $, and (b)
zero-delay time second-order correlation function $g_{2}(0)$ as
functions of the frequency detuning $\Delta _{d}/(2\protect\pi )$
of the driving force.} \label{fig4}
\end{figure}

In Fig.~\ref{fig4}, we plot the mean phonon number $\langle
n\rangle $ and $g_{2}(0)$ versus the mechanical drive detuning
$\Delta _{d}$. For a driving force tuned resonantly with the
mechanical mode frequency, we observe $g_{2}(0)\simeq $
\negthinspace 0.06. Correspondingly, the average phonon number
$\langle n\rangle \approx $ \negthinspace 0.44 (according to
Eq.~(\ref{eq9}), the maximum phonon number of the steady state in
the PB system is $\langle n\rangle _{\max }\simeq $ \negthinspace
0.5 assuming $2\sqrt{2}\epsilon \gg \kappa ).$ Thus, the system
can work as an efficient single-phonon source device with a large
output of phonons. When increasing the detuning $\Delta _{d}$,
$g_{2}(0)$ rises, while the mean phonon number $\langle n\rangle $
decreases, and around $\Delta _{d}\simeq \pm \sqrt{2}\lambda /2$
two small bunching peaks [$g_{2}(0)\!\sim $5] can be observed due
to resonantly driving the second excited states [see
Fig.~\ref{fig2}(c)].
\subsection{Discussions on increasing the robustness}

\begin{figure}[tbph]
\centering
\includegraphics[width=8.8cm]{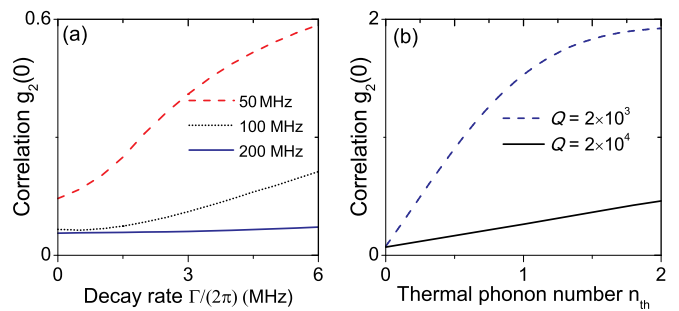}
\caption{(Color online) (a) Dependence of the sub-Poissonian dip
$g_{2}(0)$ on the decay rate of the qubit, for different values of
the driving strength $\Omega _{\text{p}}/(2\protect\pi )=50$~MHz
(red dashed curve), $\Omega _{\text{p}}/(2\protect\pi )=100$~MHz
(black dot curve) and $\Omega _{\text{p}}/(2\protect\pi )=200$~MHz
(blue solid curve)$.$ (b) $g_{2}(0)$ vs thermal phonon number
$n_{\text{th}}$ for different values of the quality factor
$Q=5\times 10^{3}$ (blue dashed curve) and $Q=5\times 10^{4}$
(black solid curve)$.$ The other parameters used here are the same
as those in Fig.~\protect\ref{fig3}(a).} \label{fig5}
\end{figure}

Equation~(\ref{eq12}) indicates that PB strongly depends on the
ratio $4\lambda ^{2}/\Gamma $. Once the relation
Eq.~(\ref{StrongCouplingCondition}) is not valid, phonon
antibunching in this hybrid system might not occur. In
Fig.~\ref{fig5}(a) we show how $g_{2}(0)$ depends on $\Gamma $ and
the nonlinear coupling strength $\lambda $ (during the numerical
simulations, $\lambda $ is adjusted by changing the drive
strength). From the results in Fig.~\ref{fig5}(a), it can be found
that PB is preserved under the drive strength $\Omega
_{\text{p}}/(2\pi )=200$~MHz [$\lambda /(2\pi )=2.56$~MHz, the
blue solid curve in Fig.~\ref{fig5}(a)] with $g_{2}(0)\!\simeq
$0.07, even when $\Gamma /2\pi $ increases to 6~MHz. However, when
$\Omega _{\text{p}}/2\pi $ decreases to 50~MHz [$\lambda /(2\pi
)=0.64$~MHz, the red-dashed curve in Fig.~\ref{fig5}(a)],
$2\sqrt{2}\epsilon $ and $\kappa $ are comparable to $4\lambda
^{2}/\Gamma $ when $\Gamma /(2\pi )$ exceeds $\sim 3$~MHz, and
phonon antibunching is more fragile to the rapid decay rate of the
qubit. We conclude that the strong drive strength can increase the
robustness of the PB against the rapid decay of the qubit.

The thermal phonons in the environment can also destroy PB to some
extent. To beat the thermal noise, one can increase the induced
nonlinearity of this PB system, or employ the NAMR of a
high-quality factor. Here we set the quality factor $Q$ as
5$\times $10$^{3}$ and 5$\times $10$^{4}$, respectively, and plot
two curves describing $g_{2}(0)$ as a function of $n_{\text{th}}$
in Fig.~\ref{fig5}(b). For $Q$=5$\times $10$^{3}$, $g_{2}(0)$ is
almost $\sim 2$ when the thermal phonon number increases to
$n_{\text{th}}=2, $ indicating that the NAMR is described by the
super-Poissonian distribution due to the thermal noise. However,
under $Q$=5$\times $10$^{4}$, $g_{2}(0)$ rises much slower with
increasing $n_{\text{th}}$. To observe better PB without being
destroyed by the thermal phonons, we can employ a high-quality
factor NAMR to decouple it from the thermal environment.

\begin{figure}[tbph]
\centering
\includegraphics[width=8.4cm]{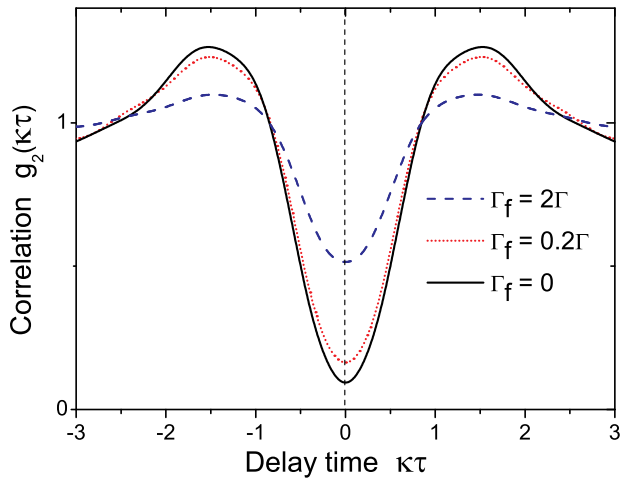}
\caption{(Color online) Steady-state second-order correlation
function $g_{2} $ of the mechanical mode versus the rescaled delay
time $\protect\kappa \protect\tau ,$ assuming various values of
the pure-dephasing rate $\Gamma _{f}=0$ (black solid curve),
$\Gamma _{f}=0.2\Gamma $ (red dot curve), $\Gamma _{f}=2\Gamma
$(blue dashed curve), where $\Gamma $ is the qubit decay rate and
$\protect\kappa $ is the decay rate of the NAMR. Here the other
parameters are the same as those in Fig.~\protect\ref{fig3}(a).}
\label{fig6}
\end{figure}

So far we have not considered the effect of pure dephasing of the
qubit. In Fig.~\ref{fig6}, we show the second-order correlation
function $g_{2}(\tau )$ under different values of the pure
dephasing rate $\Gamma _{f}$. We find that the rapid dephasing
rate will destroy the blockade effects; but even when $\Gamma
_{f}$ =2$\Gamma ,$ $g_{2}(0)$ is still about 0.5, indicating that
the NAMR still exhibits phonon antibunching. In experimental
implementations, to minimize the dephasing noise, the qubit should
be operated around its degeneracy point, and the dephasing time
has been measured even longer than 500~ns [i.e., $\Gamma
_{f}/(2\pi )\sim 0.3$~MHz] as reported in Ref.~\cite{Vion02}.
Alternatively, by coupling the mechanical mode to the SQUID loop,
we can also obtain a strong coupling of a linear longitudinal
form, which has been demonstrated in Refs.~\cite{Xue07,Zhou06}.
With the tunneling energy being in the dominant position and at
the degeneracy point, the dephasing due to charge fluctuations can
also be minimized. Another approach is to improve the properties
of the Josephson junctions and materials to eliminate excess
sources of $1/f$ noise~\cite{Koch07,Martinis05}. All these
strategies will increase the dephasing time significantly, and the
effect of pure dephasing of the qubit can effectively be
suppressed.

\section{Detecting phonon blockade by measuring the qubit}

Another challenge for phonon blockade is its detection. In
Ref.~\cite{Liu10}, the authors demonstrated that, in principle, PB
could be measured via the power spectrum of the induced
electromotive force between the two ends of the NAMRs: the
observation of extra peaks in the power spectrum means the
deterioration of PB. Another detecting method has been shown in
Ref.~\cite{Didier11}: With the NAMR resonantly coupled with a
microwave resonator cavity, the photons in the cavity will be
entangled with phonons, and they will share the same dynamics. If
detections of photons indicate photon blockade, we can conclude
that the NAMR is in a PB state. In this induced second-order
nonlinear system, the detection might be easier: energy exchanging
via the nonlinear term makes it possible to obtain the information
of the NAMR by detecting the qubit.

\begin{figure}[tbph]
\centering
\includegraphics[width=8.8cm]{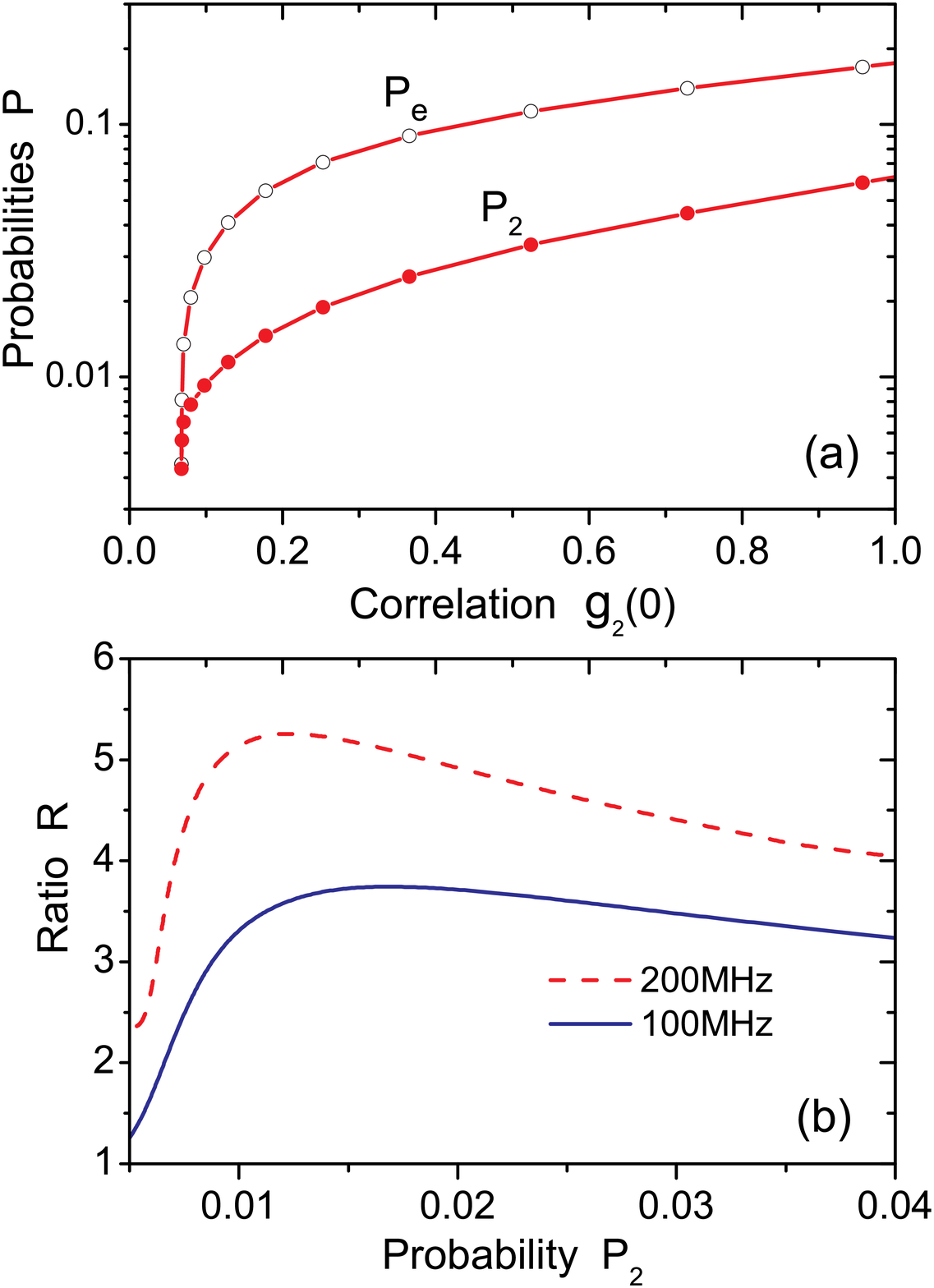}
\caption{(Color online) (a) Probabilities $P_{e}\simeq
|C_{0e}|^{2}$ and $P_{2}\simeq |C_{2g}|^{2}\,$, as defined in
Eq.~(\protect\ref{eq5}), as a function the zero-delay-time
second-order correlation function $g_{2}(0)$ under the driving
strength $\Omega _{\text{p}}/(2\protect\pi )=100$~MHz. (b) Ratio
$R=P_{e}$/$P_{2}$ versus $P_{2}.$ Here we set $\Omega
_{\text{p}}/(2\protect\pi )$ equals to 100~MHz (blue solid curve)
and 200~MHz (red dashed curve). The other parameters used here are
the same as those in Fig.~\protect\ref{fig3}(a).} \label{fig7}
\end{figure}

For single-PB systems, the most important signature is the small
probability of the incoming second phonon. Under low-power driving
for the qubit and NAMR, the probabilities $P_{e}$ and $P_{2}\
$(denoting the qubit in its excited state and the NAMR in the Fock
state $|2\rangle $), are approximately equal to $P_{0e}$ and
$P_{2g}$, respectively; that is, $P_{e}\simeq P_{0e}$ and
$P_{2}\simeq P_{2g}.$ Due to the effective nonlinear coupling
between the NAMR and the qubit, there is a coherent transition
$|e,0\rangle \leftrightarrow |g,2\rangle $ and the relation
between $P_{0e}$ and $P_{2g}$ of the steady states under the weak
driving $\epsilon \ll \lambda $ is given in Eq.~(\ref{eq8a}):
$P_{2g}$ is proportional to $P_{0e}.$ Thus, we can measure $P_{e}$
to estimate the population of the second phonon being excited, and
the sensitivity of this detection is decided by the amount
($\lambda /\Gamma )^{2}$: for larger ($\lambda /\Gamma )^{2}$, the
smaller $|C_{2g}|^{2}$ will lead to a larger probability of the
qubit in its excited state. In Fig.~\ref{fig7}, we plot both
$P_{e}$ and $P_{2}$ as a function of $g_{2}(0),$ and these
quantities are changed by increasing the drive $\epsilon $. From
Fig.~\ref{fig7}(a), it can be found that when strong
sub-Poissonian phonon statistics is observed [$g_{2}(0)\simeq
0.07$], both $P_{2}$ and $P_{e}$ are of extremely low amplitude.
When $g_{2}(0)\ $starts to rise, $P_{2}$ and $P_{e}$ increase very
rapidly. Thus, the observation of the qubit in its excited state
can be a signature of imperfect single-PB, and a higher
probability $P_{e}$ indicates a worse phonon blockade effect for
the NAMR.

In Fig.~\ref{fig7}(b), we set the driving strength of the qubit
$\Omega _{\text{p}}/(2\pi )$ equal to 100 MHz and 200 MHz,
respectively, and define the ratio $R=P_{e}$/$P_{2}\ $to estimate
the sensitivity: a large ratio $R$ means that this detection is
more sensitive. We find that for a stronger drive, $R$ is always
much higher than that of the weaker drive case. We conclude that
the large effective coupling strength does not only benefit the
robustness against environmental noise, but also increases the
sensitivity of the PB measurement.

\section{Conclusion}

We have shown how to observe phonon blockade induced by the
effective nonlinear coupling between a charge qubit and an NAMR.
This coupling could be realized in the hybrid system shown in
Fig.~\ref{fig1}. In this composite system, phonon blockade effects
will occur under resonant driving for the NAMR. By analyzing the
solution of the system steady states, we have obtained the
conditions for strong phonon antibunching and sub-Poissonian
phonon statistics. Specifically, the ratio $4\lambda ^{2}/\Gamma $
must exceed the phonon decay rate $\kappa $\ and the driving force
strength $\epsilon .$

In the numerical section, we have discussed how to more
efficiently observe phonon blockade in our proposal: (1) A
relatively strong nonlinear coupling $\lambda $ should be induced,
which can increase the robustness of phonon blockade against
different types of noise. This could be realized by increasing the
strength of the driving field (or the longitudinal coupling
strength); but both driving and coupling strengths should be
controlled within certain regimes to avoid the rapid oscillating
terms from destroying the blockade. (2) The dephasing noise of the
qubit should be suppressed, and the temperature should be low
enough so that thermal phonons are negligible. For a NAMR
oscillating at several GHz, the thermal occupation number will be
about 10$^{-3}$ at temperatures $\sim $10~mK, which is within the
capability of dilution refrigerators. (3) The quality factor $Q$
of the NAMR should be high enough to guarantee that the mechanical
mode decouples from the thermal environment.

Moreover, we have shown how to use the qubit as a detector to
check the imperfections of PB. The numerical results indicate that
the sensitivity of this detection can benefit from the strong
nonlinear coupling between the NAMR and the qubit. Besides
engineering the NAMR into PB states, the induced second-order
nonlinearity in our proposal can also be used to demonstrate some
other quantum effects of mechanical motions, such as squeezing and
superposition states (Schr\"{o}dinger cat-like
states)~\cite{Tan13,Zhou06}. All parameters in our proposal are
within experimentally accessible regimes, so it might be an
efficient method to observe quantum features of nanomechanical
motions.

\section*{Acknowledgments}

X.W. is supported by the China Scholarship Council (Grant No. 201506280142). F.N. is
partially supported by the RIKEN iTHES Project, the IMPACT program
of JST, and a Grant-in-Aid for Scientific Research (A). We acknowledge the support of a grant from the John Templeton Foundation.

\end{document}